\documentclass[preprintnumbers, floatfix, 11pt, letterpaper, superscriptaddress,nofootinbib]{revtex4}%\pdfoutput=1
\usepackage[dvipdfmx]{graphicx}
\usepackage{microtype}
\usepackage{amsmath}
\usepackage{amssymb}
\usepackage{subfigure}
\usepackage{float,color} 
\usepackage{url}
\usepackage{xcolor}
\usepackage{color}
\usepackage{mathrsfs}
\usepackage{amsfonts}
\usepackage{latexsym}
\usepackage{ragged2e}
\usepackage{epsfig}
\usepackage{textcomp}
\usepackage{phaistos}

\makeatletter
\renewcommand\@makefnmark{\hbox{\@textsuperscript{\normalfont\color{purple}\@thefnmark}}}
\renewcommand\@makefntext[1]{%
\parindent 1em\noindent
\hb@xt@1.8em{%
	\hss\@textsuperscript{\normalfont\@thefnmark}}#1}
\makeatother

\usepackage{caption}
\DeclareCaptionJustification{justified}{\leftskip=0pt \rightskip=0pt \parfillskip=0pt plus 1fil}
\def\beq{\begin{equation}}
\def\eeq{\end{equation}}

\def\mathbb{\Bbb}

\newcommand{\alf}{Alfv\'en\ }

\newcommand{\al}{\alpha}
\newcommand{\be}{\beta}

\newcommand{\vp}{\varphi}

\newcommand{\pa}{\partial}
\newcommand{\df}{\dfrac}

\definecolor{vividviolet}{rgb}{0.62, 0.0, 1.0}
\definecolor{amaranth}{rgb}{0.9, 0.17, 0.31}
\definecolor{palatinateblue}{rgb}{0.15, 0.23, 0.89}
\definecolor{brightpink}{rgb}{1.0, 0.0, 0.5}
\definecolor{cornflowerblue}{rgb}{0.39, 0.58, 0.93}
\definecolor{deepcarminepink}{rgb}{0.94, 0.19, 0.22}
\definecolor{radicalred}{rgb}{1.0, 0.21, 0.37}
\colorlet{RED}{red}

\graphicspath{{Images/}}

\graphicspath{{Images/}}

\makeatletter
\def\@fnsymbol#1{\ensuremath{\ifcase#1
	\or $\textleaf$ \or $\PHplaneTree$ \or $\PHrosette$ \or $\PHvine$
	\else\@ctrerr\fi}}%

\makeatother
\def\sideremark#1{\ifvmode\leavevmode\fi\vadjust{\vbox to0pt{\vss% the remark
		\hbox to 0pt{\hskip\hsize\hskip1em%                          will appear only
			\vbox{\hsize1.5cm\tiny\raggedright\pretolerance10000%          on the side
				\noindent #1\hfill}\hss}\vbox to8pt{\vfil}\vss}}}%
\begin{document} 
\title{Hamiltonian Formulation of Relativistic Magnetohydrodynamic Accretion on a General Spherically Symmetric and Static Black Hole: Quantum effects on Shock States}

\author{Mustapha Azreg-A\"inou}
\email{azreg@baskent.edu.tr}
\affiliation{Ba\c{s}kent University, Engineering Faculty, Ba\u{g}lica Campus, 06790-Ankara, Turkey}

\author{Mubasher Jamil}
\email{mjamil@sns.nust.edu.pk (corresponding author)}
\affiliation{School of Natural Sciences, National University of Sciences and Technology, H-12, Islamabad, 44000, Pakistan}

\author{Sousuke Noda}
\email{snoda@cc.miyakonojo-nct.ac.jp}
\affiliation{National Institute of Technology, Miyakonojo College, Miyakonojo 885-8567, Japan}

\begin{abstract}
	{\bf Abstract:} In this paper, our aim is to extend our earlier work [A. K. Ahmed et al., Eur. Phys. J. C (2016) 76:280] thereby investigating an axisymmetric plasma flow with the angular momentum onto a spherical black hole. 
	To accomplish that goal, we focus on the case that the ideal magnetohydrodynamic approximation is valid and utilizing certain conservation laws which arise from certain symmetries of the system.  After formulating a Hamiltonian of the physical system, 
we solve the Hamilton equations and look for critical solutions of (both in and out) flows. Reflecting the difference from 
	the Schwarzschild spacetime, the positions of sonic points (fast magnetosonic point, slow magnetosonic point, Alfv\'en point) 
	are altered. We explore several kinds of
flows including critical, non-critical, global, magnetically arrested and shock induced. Lastly we analyze the
shock states near a specific quantum corrected Schwarzschild black hole and deduce that quantum effects do not favor shock states by pushing the shock location outward.
\end{abstract}
%%%%%%%%
%\pacs{04.70.Bw, 04.20.-q, 04.30.Nk, 52.35.Bj} 
%\keywords{Blandford-Znajek solution, superradiance for \alf waves, force-free electromagnetism}

\maketitle
\newpage

\section{Introduction}

Black holes are one of the enduring predictions of the Einstein’s theory of general relativity (GR). For more than fifty years after GR was proposed, black holes (BHs) remained a mathematical curiosity which lead theorists to find several new BH solutions by solving the Einstein’s field equations and a few basic gravitational collapse models \cite{Oppenheimer:1939ue}. However in the 70’s and 80’s, there first appeared empirical evidences for the existence of stellar mass BHs with the observations of Cygnus X-1 binary system \cite{Webster1972,1972Natur.235..271B,1986ApJ...308..110M}. Further the detection and controversy of mysterious quasars in the 60’s lead the idea of supermassive black holes due to their ultracompactness and superluminous characteristics \cite{1963Natur.197.1040S}. In the last few years, more direct evidences in the favor of existence of both intermediate and supermassive BHs have been gathered using the observations of gravitational waves by the merger of BHs by LIGO/Virgo collaboration \cite{LIGOScientific:2016aoc}, as well as the detection of shadows of supermassive BHs at the center of M87 \cite{EventHorizonTelescope:2019ths} and Milky Way \cite{EventHorizonTelescope:2022wkp} galaxies by the EHT collaboration. These discoveries have led to the rise of phenomenological studies of testing black hole solutions in different modified gravity theories for constraining the arbitrary parameters of the spacetime metric and the underlying theory.

Historically, the simplest model of accretion of matter on a stellar body was studied by Bondi, mainly considering the accretion flow governing by Newtonian laws \cite{1952MNRAS.112..195B}. Later this model was extended by Michel considering the steady state hydrodynamical accretion with the fluid having polytropic equation of state. In this model, the fluid flow admits transonic solution \cite{1972Ap&SS..15..153M}. Based on the Michel’s model of spherical and steady state accretion, numerous extensions have been made by other authors by considering axisymmetric flows as well as accretion of exotic matter which includes dark matter and dark energy \cite{Nampalliwar:2021tyz,Bahamonde:2015uwa,Jamil:2008bc}. In literature, numerous models of accretion on black holes exist, for a review see \cite{Abramowicz:2011xu}. We proposed a model of spherical accretion of an ideal fluid for spherically symmetric black holes modeling the system as a dynamical system in a plane $(r,v)$, where $r$ is the radial coordinate and $v$ is the three dimensional speed of the fluid \cite{Ahmed:2015tyi} which has been further explored by other authors as well \cite{Yang2015,Azreg2018,Farooq2020,Yang2021,Ahmed:2016cuy}. By solving the Hamilton equations, we determined the sonic points (also known as critical points) and non-sonic critical points for ordinary fluids and fluids with exotic equation of states. We explored the implications of isothermal and polytropic fluids for spherical accretion. In this article, we are interested to extend our previous work for the magnetohydrodynamic inflow towards the spherically symmetric black holes.

The motivation for pursuing the present investigation arises from an exciting discovery made by the Event Horizon Telescope. In 2021, the EHT team imaged polarized emission around the supermassive BH in M87 on event horizon scales. This synchrotron polarized emission probed the underlying structure of the magnetic fields and plasma near the BH. The EHT team reported the average number density $n_e\sim10^{4-7}\text{cm}^{-3}$, magnetic field strength $B\sim1-30\text{G}$, and electron temperature $T_e\sim10^{10}\text{K}$ \cite{EventHorizonTelescope:2021srq}. Their model predicted that the M87 central black hole has a mass accretion rate of $(3-20)\times10^{-4}M_\odot\text{yr}^{-1}$. This observation clearly suggests that the M87* BH accretes magnetized plasma whose dynamics are governed by both gravitational and magnetic fields. It is no surprise though that large supermassive black holes like M87* have so small magnetic fields and vice versa, due to the inverse relationship between the BH gravitational radius and the magnetic field strength \cite{Chakraborty:2022itc}. As a consequence, BHs larger than $10^{10}$ solar mass are undetectable and remain in obscurity. Separate studies in literature dealing with BH shadows explore the effects of magnetized and non-magnetized plasmas on the geometry of the BH shadow \cite{Atamurotov:2021cgh,Pahlavon:2024caj,Hoshimov:2023tlz}. These studies reveal that the shadow radius alters due to plasma effect and may increase or decrease by adjusting model free parameters. In addition, it has been noted that strong magnetized plasma regions near the black holes are formed as a result of shocks in MHD plasma \cite{Fukumura:2016guw}, hence shocks generation should be investigated alongside the accretion mechanism of black holes. 

In literature, several models of general relativistic MHD accretion for black holes are investigated with the assumptions including ideal MHD, advective and viscous \cite{Mitra:2022iiv,Foucart:2015cws}. The MHD flow on 
to a Schwarzschild black hole is discussed in \cite{Mobarry1986} by solving Grad-Shafranov equation. This equation describes the interaction between the frozen-in plasma and the surrounding global magnetic field. For Kerr case, to solve Grad-Shafranov equation is much more difficult, therefore the poloidal structure of magnetosphere, which corresponds to the shape of magnetic surface (equivalently given stream function $\Psi(r,\theta)$) is given by hand. In \cite{Takahashi1990}, the authors assumed 
the stream function, and fluid flow along the stream line is investigated. In this paper, instead of assuming the shape of the magnetic surface, we consider an axisymmetric inflow restricted in a plane around the black hole by fixing the value of an angular coordinate, which is similar analysis to inflow on the equatorial plane of a Kerr black hole \cite{Gammie1999}.

This paper is structured as follows: In Sec. II, we present the governing equations of accretion of plasma on a general static and spherical black hole, along with the symmetries of the model and the corresponding conservation laws. Sec. III deals with the solution of the governing equations utilizing the conservation laws. In particular, we determine the Hamiltonian of our system, determine the fluid velocity components, solve the Hamilton equations to finding the critical points as well. As a special case, we analyze the accretion model of the polytropic fluid. In Sec. IV, we study a special model of spherical MHD accretion on a Schwarzschild black hole and present the conditions under which inflow or outflow can occur in the vicinity of black hole. Here we explore several kinds of flows including critical, global, magnetically arrested and shocks in flows. Lastly we analyze the shock states near the quantum corrected Schwarzschild black hole.  Throughout we work in the units $c=G_N=\hbar=1,$ unless mentioned otherwise.

\section{Magnetohydrodynamics in the spherical black hole spacetime}

As the background spacetime, we consider a static and spherical black hole with (or without) quantum correction, of which the general form of metric is given as
%line element $ds^2=g_{\lambda \nu}dx^{\lambda}dx^{\nu}$ is given as 
%%%
\begin{equation}
	ds^2=g_{tt}(r)dt^2+g_{rr}(r)dr^2 + g_{\theta\theta}(r) (d\theta^2+\sin^2\theta d\varphi^2).
	\label{metric}
\end{equation}
%%%
where the components of the metric
%$\al(r), \be(r)$, and $\ga(r)$ 
are functions of the coordinate $r$, which depends on gravitational theories, quantum correction, and so on. Here we do not assume the functional form yet, but when deriving MHD flow solutions, we shall consider one kind of spherical black hole.

Now, we introduce the magnetohydrodynamics in curved spacetime, {endowed with spherical symmetry} within the ideal MHD approximation that the conductivity of the fluid tends to infinity implying a vanishing electric field for comoving observer: $F_{\mu\nu}u^{\nu}= F^{\mu\nu}u_{\nu} = 0$, where $u^{\mu}$ is the four-velocity of flow satisfying $u_\mu u^\mu=-1$. We apply the method introduced in the following pioneering papers \cite{Bekenstein1978, Camenzind1986a,Camenzind1986b,Camenzind1987,Takahashi1990,Takahashi2006}. Under these assumptions, we investigate the ideal MHD steady and advective inflow onto 
a spherically symmetric black hole. The MHD flow is governed by the following three conserved laws (namely, the particle number conservation, the energy conservation, the angular momentum conservation), the Maxwell equation and the ideal MHD condition, respectively \cite{Abramowicz:2011xu}
%and \cite{,Fendt1996,Fendt2001,Fendt2004}. 
% In this work we will restrict ourselves to solutions endowed with spherical symmetry and in a subsequent paper we will consider axisymmetric solutions. 
% The study of an MHD inflow onto a rotating black hole was discussed in~\cite{Takahashi1990}, and in the recent papers~\cite{Pu2015,Mitra2022}. In these seminal works, the authors solved special or general relativistic MHD equations to describe the inflow and outflow motions of the magnetohydrodynamic fluid in the Kerr spacetime.
\begin{align}
	\label{baryon} &\nabla_\mu(n u^{\mu})=0,\\
	\label{Econsv} &\nabla_\mu (-T^{\mu}_{\ \ \nu}k^\nu)=0,\\
	\label{Lconsv} &\nabla_\mu (T^{\mu}_{\ \ \nu}m^\nu)=0,\\
	\label{SF} &\nabla_\mu{\ }^{*}F^{\mu \nu}  = 0,\\
	\label{IdealMHDcond} & F^{\mu \nu} u_{\nu} = 0,
\end{align}
where $n$ is the proper particle number density, $k^\mu$ and $m^\mu$ are Killing vectors: $k^\mu=(1,0,0,0)$ and $m^\mu=(0,0,0,1)$, $^{*}F^{\mu \nu}$ is the dual to $F^{\mu \nu}$, and $T^{\mu\nu}$ is the energy momentum tensor
\beq
T^{\mu\nu}=(\rho+p) u^{\mu} u^{\nu}+p g^{\mu \nu}+\dfrac{1}{4\pi} \left(F^{\mu \al}F^{\nu}_{\ \al}-\dfrac{1}{4}g^{\mu \nu} F^{\al \be}F_{\al\be}\right),
\eeq
which consists of the plasma part with the total energy density of plasma gas $\rho$ and its pressure $p$, and electromagnetic field part.
\vskip4pt

\begin{figure}[H]
	\centering
	\includegraphics[width=0.65\linewidth]{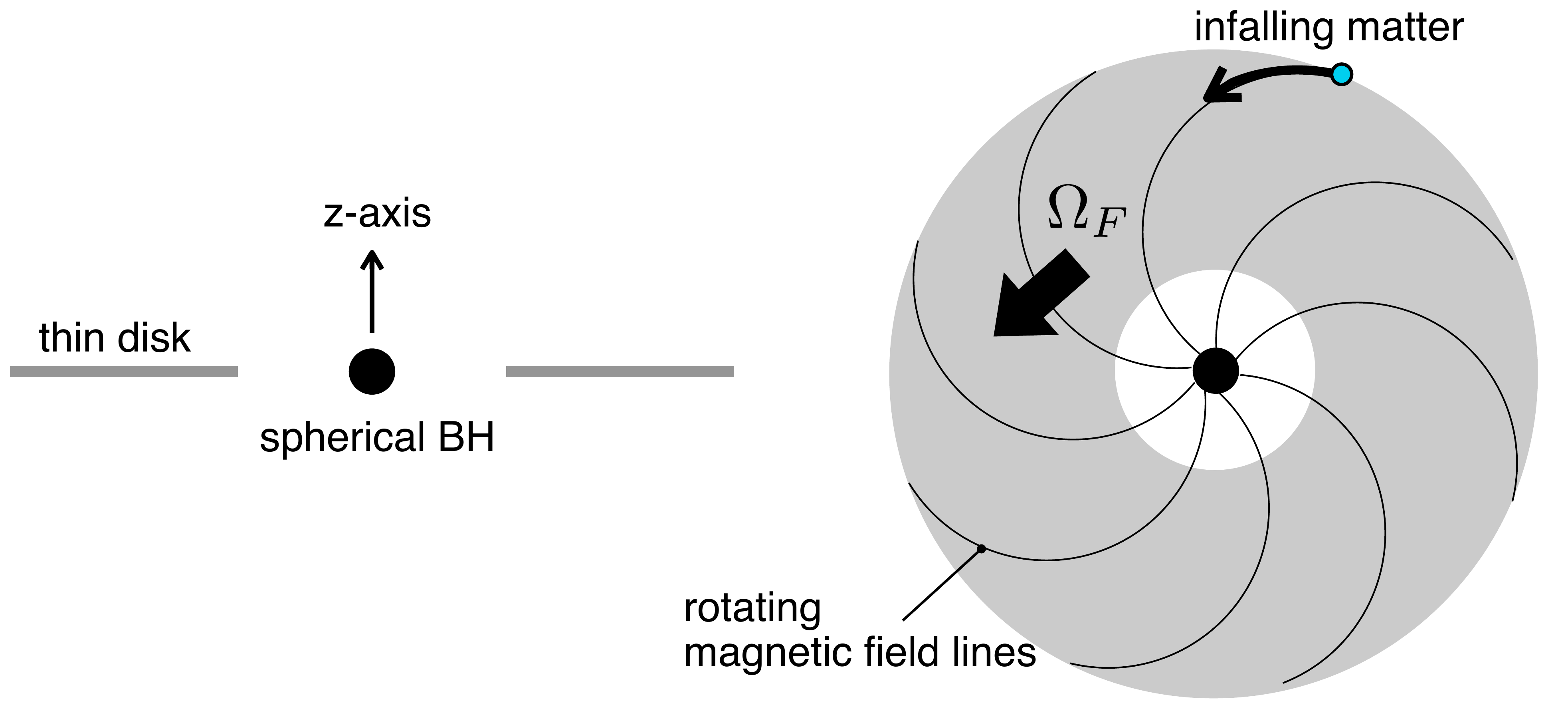}
	\caption{\footnotesize{Schematic of an infalling plasma: The left and right panels depict the views of the present inflow from the edge-on direction and from z-axis, respectively. The matter starts from the outer edge of the thin accretion disk.}}
	\label{fig:schematic}
\end{figure}

From now on we assume that the plasma is confined in a narrow (thin) accretion disk located in the equatorial plane $\theta =\pi/2$ in such a way that the component $u^\theta$ of the velocity vector is taken to be null (i.e., the matter from the accretion disk does not leave the equatorial plane throughout the motion). As we shall see below this implies $B^\theta=0$. The plasma is distributed symmetrically about the black hole (hereafter: BH) and the inflow is assumed to be axisymmetric and stationary. Under this assumption, the electromagnetic (EM) field acquires axial symmetry about the z-axis around which the plasma is revolving. In this paper, we investigate two different outer 
boundary conditions separately: The first one concerns with an in-falling matter starting its motion from the outer edge $r=r_\text{edge}$ on the equatorial plane of the BH with negligible radial velocity as shown in Fig.~\ref{fig:schematic}, while the second boundary is the case of accretion of the wind from a companion object, for which the radial flow velocity at the outer edge may be so fast that it is supersonic initially.
% The inflowing matter starts its journey from the outer edge $r=r_\text{edge}$ on the equatorial plane of the BH with negligible radial velocity as shown in Fig.~\ref{fig:schematic}.
\vskip4pt  
%%%
\noindent
For stationary and axisymmetric EM  field, the Lie derivative of the EM field tensor is zero: 
${\cal{L}}_\text{k}F_{\al \be}=0, \ {\cal{L}}_\text{m}F_{\al \be}=0$, where $\text k=k^\mu\partial_\mu$ and $\text m=m^\mu\partial_\mu$ are Killing vectors. In the coordinates $(t,r,\theta,\vp)$ describing the metric, the Lie derivatives for the EM field tensor become respectively
%%%
\beq
F_{\al \be,t}=0,\ \ F_{\al \be,\varphi}=0.
\label{eq:Ft}
\eeq
%%%
% Likewise, other quantities satisfy
% %%%
% \beq
% {\cal{L}}_{\xi} u^{\mu}=0,\ \ {\cal{L}}_{\xi} n=0,\ \ {\cal{L}}_{\xi} \rho=0.
% \eeq
% %%%
One of the Maxwell equations $F_{[\al \be;\gamma]}=F_{[\al \be,\gamma]}=0,$ and \eqref{eq:Ft} yield 
%%%
\beq
F_{t \varphi,r}=0,\ \ \ F_{t\varphi ,\theta}=0.
\label{Ftvpderiv}
\eeq
%%%
Therefore, $t\vp$-component of the field strength satisfies 
$F_{t\vp ,\mu}=0$, and equivalently $F_{t\vp}=\text{const.}$ We can choose this constant to be zero by imposing some appropriate asymptotic conditions. Hence,
%%%
\beq
F_{t \vp}=0.
\label{Ftvp}
\eeq
Using this along with~\eqref{IdealMHDcond} and the symmetry constraints $u^\theta=0$ \footnote{The assumption $u^\theta =0$ has an important consequence: It implies that any quantity that is conserved on each flow line is conserved for all $r$. In fact, since partial derivatives of all fields with respect to $t$ and $\varphi$ are zero, the relation $d(.)/d\tau =0$, where $(.)$ represents any quantity that is conserved on each flow line as $\Omega_F$ and $\eta$,  implies $d(.)/d\tau =u^\alpha (.)_{,\alpha}=u^r(.)_{,r}=0$, and thus $(.)_{,r}=0$.} and the fact that no EM field depends on ($t,\, \varphi$) coordinates, then the Maxwell equations $F_{[\al \be;\gamma]}=F_{[\al \be,\gamma]}=0$, yield $F_{tr}=F_{\varphi r}=0$. For stationary and axisymmetric MHD flow, these same symmetry constraints yield the following conservation laws where $\Omega_F$ and $\eta$ are constants along each flow line (i.e., $d\Omega_F/d\tau =0$ and $d\eta/d\tau =0$) \cite{Bekenstein1978}:
\begin{equation}\label{m1}
	\Omega_F := -\frac{F_{t\theta}}{F_{\varphi\theta}}, \qquad
	\frac{F_{\varphi\theta}}{\sqrt{-g}nu^r} := -\frac{1}{\eta},\qquad
	\frac{F_{r\theta}}{\sqrt{-g}n(u^\varphi + \Omega_F u^t)}=\frac{1}{\eta},
\end{equation}
%where ($\Omega_F,\,\eta$) are constants along each flow line. 
Note that $F_{tr}=F_{\varphi r}=0$ because of the constraint $u^\theta=0$; this is to say that in the limit $u^\theta\to 0$, we have $F_{tr}\to 0$ and $F_{\varphi r}\to 0$ but $F_{tr}/F_{\varphi r}=\Omega_F$~\cite{Bekenstein1978}. On combining the equations given in~\eqref{m1}, we obtain
\begin{equation}\label{m2}
	F_{r\theta}\,\dot{r} = F_{\vp \theta} (\Omega_F - \Omega),
\end{equation}
which can be derived directly from the ideal MHD condition~\eqref{IdealMHDcond}. Here $\dot{r}\equiv dr/dt$, $\Omega \equiv u^\vp /u^t=\dot{\varphi}$ is the angular velocity of the plasma flow, and $\Omega_F$ is the angular frequency of the streamline which is constant along each flow line, manifesting the well known line isorotation law.\vskip4pt

The mass conservation law \eqref{baryon} becomes $\pa_r (\sqrt{-g}n u^r)=0$ due to the symmetry assumptions. This implies  
\begin{equation}\label{C1}
	\sqrt{-g}n u^r=\text{const}=C\,,\quad \text{for all }r\,,
\end{equation}
where $C$ denotes the BH accretion rate.

Using the above conserved quantities, the field strength can be written in terms of the flow velocities as 
%%%
\beq
F_{\mu\nu}
=\begin{pmatrix}
	0 & 0  & F_{t \theta }  & 0 \\  
	&0 & F_{r\theta} & 0 \\ & &0 &F_{ \theta \vp} \\ & & &0
\end{pmatrix}
=\df{\sqrt{-g}\ n}{\eta}
\begin{pmatrix}
	0 & 0  & \Omega_F u^r  & 0 \\  
	&0 &u^\vp +\Omega_F u^t & 0 \\ & &0 &u^r \\ & & &0
\end{pmatrix}.
\label{eq:Fmn}
\eeq
%%%

The expression of the magnetic field, measured in a laboratory frame, is defined by $B_\mu =-\eta_{\mu\nu\alpha\beta}k^\nu F^{\alpha\beta}/2$, where $\eta_{\mu\nu\alpha\beta}=\sqrt{-g}\,\epsilon_{\mu\nu\alpha\beta}$ ($\epsilon_{tr\theta\varphi}=+1$). As we noticed earlier, $F_{tr}=F_{\varphi r}=0,$ because of the constraint $u^\theta=0$, now inserting $F_{\varphi r}=0$, into the expression of $B_\mu$ yields $B_\theta=B^\theta=0$. Introducing the poloidal magnetic field $B$ and the poloidal velocity of fluid $u$ by $B^2 :=(B_rB^r + B_\theta B^\theta)/g_{tt}^2=-(F_{\theta\varphi})^2/(g_{tt}g_{\varphi\varphi}^2)$ and $u^2 := u_ru^r + u_\theta u^\theta =g_{rr}(u^r)^2$, we see that $\eta$ given in \eqref{m1} reduces to 
%\begin{align}\label{m4}
%	&B^2 := \frac{B_rB^r + B_\theta B^\theta}{g_{tt}^2}=\frac{B_r^2}{g_{tt}^2g_{rr}}=\frac{|g|(F^{\theta\varphi})^2}{g_{tt}^2g_{rr}}=-\frac{(F_{\theta\varphi})^2}{g_{tt}g_{\varphi\varphi}^2},\\
%	&u^2 := u_ru^r + u_\theta u^\theta =g_{rr}(u^r)^2,
%\end{align}
\begin{equation}\label{m5}
	|\eta| =\bigg|\df{\sqrt{-g}\,nu^r}{F_{\theta \vp}}\bigg| = \bigg|\sqrt{-g_{tt}g_{rr}}\,g_{\varphi\varphi}\,\frac{nu^r}{F_{\theta \vp}}\bigg| =\bigg| n\frac{ u}{B}\bigg|,
\end{equation}
and represents the particle-flux per unit flux-tube~\cite{Takahashi1990} or the mass flux per unit poloidal magnetic field.\vskip4pt

%\textcolor{red}{MJ: For the sake of clarification, it would be suitable to express all the $F_{\alpha\beta}$ in terms of electric and magnetic field components.}

%\textcolor{red}{MJ: The order of the sections seems random. The sections should start with general spacetime, underlying plasma energy-momentum tensors, symmetries and Killing vectors and then conservation laws. \\
	%Most importantly, we should mention what underlying problem is going to be addressed and how.}

% We eliminate $F_{t\theta}$ using \eqref{eq:rel1} and \eqref{eq:rel2}, and obtain
% %%%
% \begin{align}
	% F_{tr;\theta}+\left(\dfrac{u^r}{u^\theta}F_{tr}\right)_{;r}=0 \ \ \Rightarrow \ \ -u^\theta \left(\dfrac{u^r}{u^\theta}\right)_{,r}=\dfrac{d}{d\tau}\log{F_{tr}}.
	% \label{eq:rel3}
	% \end{align}
% %%%
% Likewise eliminating $F_{\theta\vp}$, we get
% %%%
% \begin{align}
	% F_{r\vp;\theta}+\left(\dfrac{u^r}{u^\theta}F_{\vp r}\right)_{;r}=0 \ \ \Rightarrow \ \ -u^\theta \left(\dfrac{u^r}{u^\theta}\right)_{,r}=\dfrac{d}{d\tau}\log{F_{\vp r}}.
	% \label{eq:rel4}
	% \end{align}
% %%%
% Subtracting \eqref{eq:rel4} from \eqref{eq:rel3}, we obtain $d/d\tau (F_{tr}/F_{\vp r})=0$. 
% Therefore the ratio is a constant along the flow line. 
% Actually, the constant corresponds to the angular 
% velocity of the 
% magnetic field line \cite{Takahashi1990}:
% %%%
% \beq
% \anno{  \Omega_F:=-\dfrac{F_{t\theta}}{F_{\vp \theta}}. 
	% }
% \label{OmegaF}
% \eeq
% %%%

In addition to $\Omega_F$ and $\eta$, we have also ${\cal{E}}$ and ${\cal{L}}$, which correspond to energy and angular momentum. The conservation laws \eqref{Econsv} and \eqref{Lconsv} yield
\beq
\pa_r \left(-\sqrt{-g}T^{r}_{\ \ t} \right)=0,\ \pa_r\left( \sqrt{-g}T^r_{\ \ \vp} \right)=0,
\eeq
respectively. Therefore, we can introduce the following conserved quantities:
\begin{align}
	\label{EBernou}
	&{\cal{E}}:=\df{-\sqrt{-g}T^{r}_{\ \ t}}{F_{\theta \vp}\eta}=-\df{\sqrt{-g}}{F_{\theta \vp}\eta}\left[(\rho+p)u^r u_t +\df{g^{rr}g^{\theta \theta}}{4\pi}F_{r\theta}F_{t\theta}    \right]=-h  u_t -\df{B_\vp}{4\pi \eta}\Omega_F,\\
	\label{LBernou}
	&{\cal{L}}:=\df{\sqrt{-g}T^r_{\ \ \vp}}{F_{\theta\vp}\eta}=\df{\sqrt{-g}}{F_{\theta \vp}\eta}\left[(\rho+p)u^r u_\vp +\df{g^{rr}g^{\theta \theta}}{4\pi}F_{r\theta}F_{\vp \theta}    \right]=h  u_\vp -\df{B_\vp}{4\pi \eta},
\end{align}
where $h$ is the specific enthalpy defined as
%\footnote{
	%	The conservation of the baryon number \eqref{baryon} 
	%	gives the definition of specific enthalpy. 
	%	\beq
	%	\notag 0=(nu^{\al})_{;\al}=n_{,\al}u^{\al}+nu^{\al}_ {\ ;\al}=n_{,\al}u^{\al}+n\left(\dfrac{-\rho_{,\al}u^{\al}}{\rho+p}\right),
	%	\eeq
	%	%%%
	%	hence
	%	%%%
	%	\beq
	%	\notag \dfrac{p+\rho}{n} u^\al \partial_\al n -u^\al \partial_\al \rho =0 \ \ \Rightarrow \dfrac{p+\rho}{n}\dfrac{dn}{d\tau}=\dfrac{d\rho}{d\tau},
	%	\eeq
	%	%%%
	%where $d/d\tau:= u^\al \partial_\al$. Then we define the specific enthalpy $h$
	%%%
	%\beq
	%h:=\dfrac{d\rho}{dn}=\dfrac{p+\rho}{n}.
	%\eeq
	%} 
$h:=(\rho+p)/n$. Note that since $F_{\theta \vp}\eta = -\sqrt{-g}nu^r$ given in~\eqref{m1} is conserved for all $r$, the quantities $\cal{E}$ and $\cal{L}$ are too conserved for all $r$.

There are two remaining constants of motion, $\Phi_1,\,\Phi_2$, that can be related to $C,\,\Omega_F,\,\eta$ as follows. The $t-$ and $\varphi-$ source-free equations~\eqref{SF} yield
\begin{equation}\label{PP1}
	\Phi_1 =-\sqrt{-g}(u^tb^r-u^rb^t),\qquad \Phi_2 =\sqrt{-g}(u^rb^\varphi-u^\varphi b^r),
\end{equation}
where $b^\mu$ is the magnetic field in the fluid frame
\begin{equation}\label{PP2}
	b_\mu = - \frac{1}{2}\,\eta_{\mu\nu\alpha\beta}u^\nu F^{\alpha\beta}\,.
\end{equation}
It is then straightforward to show that
\begin{equation}\label{PP3}
	\Omega_F=\frac{\Phi_2}{\Phi_1},\qquad	\eta =-\frac{C}{\Phi_1}\,.
\end{equation}

\section{flow solutions}
To obtain flow solutions, we introduce kinematic variables describing the ideal MHD flow. Due to the symmetry in the present case, the system can be represented by a Hamiltonian which is a function of two variables namely, position of fluid parcel $r$ and its velocity $V$. Moreover, using a property of the Hamiltonian, we get the so-called wind equation giving critical points (CPs).
\subsection{Kinematic variables of the flow}
Using \eqref{m2} and the definition of $\eta$, the azimuthal component of the magnetic field $B_\vp = \sqrt{-g}F^{r\theta}$ can be written in terms of $u_t$ and $u_\vp$ 
as 
\begin{align}
	B_\vp =g^{rr}g^{\theta\theta}\sqrt{-g}\df{F_{\theta\vp}}{u^r}(u^\vp -\Omega_F u^t)=-gg^{rr}g^{\theta\theta}\df{n}{\eta} (u^\vp -\Omega_F u^t)=-\df{n}{\eta}(g_{tt}u_\vp -\Omega_F g_{\vp\vp}u_t).
\end{align}
Substituting this into \eqref{EBernou} and \eqref{LBernou}, we obtain
\begin{align}
	\label{E}&{\cal{E}}=-\left( h  +\df{n}{4\pi \eta^2} \Omega_F^2 g_{\vp\vp} \right) u_t +\df{n}{4\pi \eta^2} \Omega_F g_{tt}u_\vp,\\
	\label{L}&{\cal{L}}=-\df{n}{4\pi \eta^2 } \Omega_F g_{\vp\vp}u_t + \left( h  + \df{n}{4\pi \eta^2} g_{tt}\right) u_\vp,
\end{align}
% Solving these equations for $u_t$ and $u_\vp$, and introducing the general relativistic radial \alfv Mach number \cite{Weber1967,Takahashi1990}
% \beq
% M_\text{A}^2:=\df{4\pi h \eta^2}{n},
% \eeq
% we obtain
% \begin{align}
% 	\label{hut}
% 	&h u_t=-\df{M_\text{A}^2{\cal{E}}+g_{tt}({\cal{E}}-\Omega_F{\cal{L}})}{M_\text{A}^2-\Gamma},\\
% 	\label{huvp}
% 	&h u_\vp=\df{M_\text{A}^2{\cal{L}}-g_{\vp\vp}\Omega_F({\cal{E}}-\Omega_F{\cal{L}})}{M_\text{A}^2-\Gamma},
% \end{align}
% and $B_\vp$ can be written as 
% \beq
% \label{hB}
% hB_\vp=-\df{n}{\eta}\df{M_\text{A}^2(g_{tt}{\cal{L}}+\Omega_F g_{\vp\vp}{\cal{E}})}{M_\text{A}^2-\Gamma},
% \eeq
% where $\Gamma:=-g_{tt}-\Omega_F^2 g_{\vp\vp}$, and the 
% radii satisfying $\Gamma=0$ are called \textit{light surfaces}.
% The above three expressions have denominators that vanish at the point where $M_\text{A}^2=\Gamma$, which is 
% called the \alf point \cite{Weber1967,Camenzind1986a,Takahashi1990}. For the MHD flow to be finite, the numerators of $u_t$, $u_\vp$, and $B_\vp$ need to be zero at the \alf point as well.  This regularity condition provides the relation between $\Omega_F$ 
% and ${\cal{L}}/{\cal{E}}$:
% \begin{equation}
% 	\widetilde{\cal{L}}=-\left.\df{g_{\vp\vp}}{g_{tt}}\right|_\text{A} \Omega_F,
% \end{equation}
% where $\widetilde{\cal{L}}:={\cal{L}}/{\cal{E}}$.
% Therefore, the \alf point determines the angular momentum conveyed by the MHD flow.\vskip4pt
% }

To determine the critical points (CPs) and the profile of the flow, we will not integrate the set of differential equations governing the flow; rather, we will apply a Hamiltonian approach like we used earlier in~\cite{Azreg2016,Azreg2017,Azreg2018,Ahmed:2016cuy}. In the ($\theta=\pi/2$)-plane, rewriting~\eqref{metric} as $ds^2=-(\sqrt{-g_{tt}}dt)^2+(\sqrt{g_{rr}}dr)^2 + (\sqrt{g_{\varphi\varphi}}d\varphi^2)$, we see that the 3-velocity components are
\begin{equation}\label{k1}
	v_r=\frac{\sqrt{g_{rr}}}{\sqrt{-g_{tt}}}~\dot{r},\qquad v_\varphi=\frac{\sqrt{g_{\varphi\varphi}}}{\sqrt{-g_{tt}}}~\Omega ,\qquad v^2:=v_r^2+v_\varphi^2\,.
\end{equation}
When inflow or outflow is accompanied by fluid rotation ($\Omega\neq 0$),  as we shall see below, $v_r$ is not a convenient kinematic variable. Rather, we introduce the radial 3-velocity in the corotating frame $V_r$, simply denoted by $V$. Since the corotating frame has the 4-velocity vector $U_{\text{corot}}^\alpha =(1,\,0,\,0\,,v_\varphi)/\sqrt{-g_{tt}(1-v_\varphi^2)}$, it is straightforward to show that $V$ is related to $v_r$ by
\begin{equation}\label{k2}
	V=\frac{v_r}{\sqrt{1-v_\varphi^2}}\,.
\end{equation}
Next, we intend to express the rhs of~\eqref{E} in terms of ($r,\,V$) and the constants of motion. 
%The ratio of~\eqref{huvp} and~\eqref{hut} yields
%\begin{equation}\label{k3}
%\Omega=-\frac{M_\text{A}^2{\cal{L}}-g_{\vp\vp}\Omega_F({\cal{E}}-\Omega_F{\cal{L}})}{M_\text{A}^2{\cal{E}}+g_{tt}({\cal{E}}-\Omega_F{\cal{L}})}\,.
%\end{equation}
From $u^\alpha u_\alpha=-1,$ we obtain
\begin{equation}\label{k4}
	u_t =-\frac{\sqrt{-g_{tt}}}{\sqrt{1-v^2}}\,.
\end{equation}
%Using this last equation along with~\eqref{C1} in $u^\alpha u_\alpha=-1$, we obtain
%\begin{equation}\label{k5}
%u_{\varphi} =\frac{\sqrt{-g_{tt} g_{\varphi \varphi }^2 v^2 n^2-C^2 (1-v^2)}}{ \sqrt{-g_{\text{tt}}} \sqrt{g_{\varphi \varphi
			%}}\sqrt{1-v^2}\,n}\,.
%\end{equation}
Now, solving~\eqref{L} for $u_{\varphi}$ and using~\eqref{k4}, we get
\begin{equation}\label{k5}
	u_{\varphi} =\frac{4\pi \eta^2 {\mathcal{L}}\sqrt{1-v^2}-\Omega _F \sqrt{-g_{tt}}\, g_{\varphi \varphi } n}{(4\pi \eta^2 h +g_{tt} n) \sqrt{1-v^2}}\,.
\end{equation}
All we need now is to express $v^2$ in terms of $V^2$ to express ($u_t,\,u_\varphi$) in terms of ($r,\,V$). Before doing that there is an interesting conclusion to draw as follows: From the relation
\begin{equation}
	v_r=\frac{C \sqrt{1-v^2}}{\sqrt{-g_{tt}}g_{\varphi \varphi }\,  n}=\frac{C \sqrt{1-v_r^2-v_\varphi^2}}{\sqrt{-g_{tt}}g_{\varphi \varphi }\,  n}\,,
\end{equation}
we obtain 
\begin{equation}\label{k6}
	v_r^2=\frac{C^2 (1-v_{\varphi }^2)}{C^2-g_{tt} g_{\varphi \varphi }^2 n^2},\,
\end{equation}
which yields, upon using~\eqref{k2} the following
\begin{equation}\label{k7}
	V^2=\frac{C^2 }{C^2-g_{tt} g_{\varphi \varphi }^2 n^2}<1\,.
\end{equation}
This may seem to show that the relative 3-velocity of the fluid with respect to the corotating frame does not depend explicitly on any presence of magnetic field, which is only apparently true. As we shall see below, $n$ depends explicitly on $F^2=F_{\mu\nu}F^{\mu\nu}$. It is obvious from~\eqref{k7} that $V^2\to 1$ as $g_{tt}\to 0$, which is the case if $g_{tt}(r=r_h)=0,$ where $r_h$ is the horizon of the BH.\vskip4pt 
%Since $v_\varphi\to 1$ as $r\to r_h$, we see that $v_r\to 0$ as $r\to r_h$. 

In order to express $v^2$ in terms of $V^2$, we write $v^2$ as
\begin{equation}\label{k8}
	v^2=v_r^2+v_\varphi^2=V^2(1-v_\varphi^2)+v_\varphi^2=V^2\bigg(1-\frac{(1-v^2)u_\varphi^2}{g_{\varphi\varphi}}\bigg)+\frac{(1-v^2)u_\varphi^2}{g_{\varphi\varphi}}\,,
\end{equation}
where $u_\varphi$ is given in~\eqref{k5}. Once $u_\varphi$ is substituted into~\eqref{k8}, we set $x=\sqrt{1-v^2}$, which occurs in the rhs of~\eqref{k5}, and solve for $x$ then for $v^2$ in terms of $V^2$. We obtain two roots for the equation related by $v_1^2(\mathcal{L})=v_2^2(-\mathcal{L})$. Since $\mathcal{L}$ can be given any sign, we will work with the solution $v_1^2$  by~\eqref{A1} given in Appendix.\vskip4pt

Since the rhs of~\eqref{E} contains $n$, this also can be expressed as a function ($r,\,V$) upon reversing~\eqref{k7} yielding
\begin{equation}\label{k9}
	n(r,V)=\frac{C \sqrt{1-V^2}}{\sqrt{-g_{tt}}\, g_{\varphi \varphi } V}\,.
\end{equation}
Given an equation of state expressing $h$ in terms of $n$, we can also express enthalpy $h$
%and Alfv\'en Mach number squared $M_A^2$ 
as a function of the variables ($r,\,V$). Finally, Eq.~\eqref{E} takes the form
\begin{equation}\label{ErV}
	{\mathcal{E}}(r,V)=-\bigg(h(r,V)+\frac{\Omega _F^2 g_{\varphi \varphi }(r)\,n(r,V)}{4 \pi  \eta ^2}\bigg) u_t(r,V)+\frac{\Omega _F g_{t t}(r)\,n(r,V)}{4 \pi  \eta ^2}u_{\varphi }(r,V),\,
\end{equation}
where $u_t$ and $u_\varphi$ are given by~\eqref{k4} and~\eqref{k5}, respectively, and $v^2$ is given in terms of $V^2$ in Appendix by~\eqref{A1} and~\eqref{A2}. Since ${\mathcal{E}}(r,V)$ is a global constant, it is the {\it Hamiltonian} of our system.\vskip4pt

In the same manner, we can express the components of the magnetic field  $b^\mu$ in the fluid frame in terms of ($r,\,V$) as follows
\begin{align}
	& b^t(r,V)=-\frac{[1+u^t (u_t-u_{\varphi } \Omega _F)] n}{\eta }\,,\nonumber\\
	&\label{brV}b^r(r,V)=-\frac{u^r (u_t+u_{\varphi } \Omega _F) n}{\eta }=-\frac{C (u_t+u_{\varphi } \Omega _F)}{\eta \sqrt{-g}}\,,\\
	&b^{\varphi }(r,V)=-\frac{[\Omega _F (1+u^{\varphi } u_{\varphi })+u_t (u^{\varphi }+2 u^t \Omega _F)] n}{\eta }\,.\nonumber
\end{align}

Using all expressions given in (\ref{brV}), we express $n$ explicitly depending on $F^2=F_{\mu\nu}F^{\mu\nu}$ as follows
\begin{equation}\label{n2F2}
	n^2=\frac{\eta ^2}{2}~\frac{F^2}{g_{t t}+g_{\varphi  \varphi } \Omega _F^2+\left(u_t-u_{\varphi } \Omega _F\right)^2}\,.
\end{equation}

\subsection{Wind equation}

The wind equation is obtained upon differentiating~\eqref{ErV}:
\begin{equation}\label{w1}
	d{\cal{E}}(r,V)={\cal{E}}_{,r}\,dr + {\cal{E}}_{,V}\,dV =0,
\end{equation} 
since ${\cal{E}}$ is global constant of motion. Thus,
\begin{equation}\label{w2}
	\frac{dV}{dr}=-\frac{{\cal{E}}_{,r}}{{\cal{E}}_{,V}}\,.
\end{equation}
At the CPs where $(r,\,V)=(r_c,\,V_c),$ we have
\begin{equation}\label{CP}
	{\cal{E}}_{,r}(r_c,V_c)=0,\quad\text{and}\quad {\cal{E}}_{,V}(r_c,V_c)=0\,.
\end{equation}
Using the equation, $d\ln h=a^2 d\ln n$, for the adiabatic sound speed, we obtain
\begin{align}
	\label{CP1}&\bigg[\frac{\Omega_F n}{4\pi\eta^2 hu_t}~(g_{tt}u_\varphi -\Omega_Fg_{\varphi\varphi}u_t)-a^2\bigg](\ln n)_{,r}=(\ln u_t)_{,r} + \frac{\Omega_F n}{4\pi\eta^2 hu_t}~(\Omega_Fg_{\varphi\varphi}u_t-g_{tt}u_\varphi)_{,r} \,,\\
	\label{CP2}&\bigg[\frac{\Omega_F n}{4\pi\eta^2 hu_t}~(g_{tt}u_\varphi -\Omega_Fg_{\varphi\varphi}u_t)-a^2\bigg](\ln n)_{,V}=(\ln u_t)_{,V} + \frac{\Omega_F n}{4\pi\eta^2 hu_t}~(\Omega_Fg_{\varphi\varphi}u_t-g_{tt}u_\varphi)_{,V} \,,
\end{align}
where $(.)_{,r}$ and $(.)_{,V}$ denote partial derivatives. In this Hamiltonian formalism, $r$ and $V$ are treated as independent variables, for instance, we have $(\Omega_Fg_{\varphi\varphi}u_t-g_{tt}u_\varphi)_{,V}=\Omega_Fg_{\varphi\varphi}u_{t,V}-g_{tt}u_{\varphi,V}$ and similar expression for $(\Omega_Fg_{\varphi\varphi}u_t-g_{tt}u_\varphi)_{,r}$, where $u_{t,V}$, $u_{\varphi,V}$, $u_{t,r}$, and $u_{\varphi,r}$ are evaluated using~\eqref{k4} and~\eqref{k5}. Further, $(\ln n)_{,r}$, $(\ln n)_{,V}$, $(\ln u_t)_{,r}$, and $(\ln u_t)_{,V}$ are given by
\begin{align}
	&(\ln n)_{,r}=-\frac{1}{2}~\frac{g_{tt,r}}{g_{tt}}-\frac{g_{\varphi\varphi,r}}{g_{\varphi\varphi}}\,,& & (\ln n)_{,V}=-\frac{1}{V(1-V^2)}\,,\\
	&(\ln u_t)_{,r}=\frac{1}{2}~\frac{g_{tt,r}}{g_{tt}}+\frac{v}{1-v^2}~v_{,r}\,,& &(\ln u_t)_{,V}=\frac{v}{1-v^2}~v_{,V}\,,
\end{align}
where $v_{,r}$ and $v_{,V}$ are evaluated using the expression of $v(r,V)$ given in Appendix by~\eqref{A1} and~\eqref{A2}. Eq.~\eqref{CP2} reduces to
\begin{equation}\label{CP3}
	a^2 = \frac{vV(1-V^2)}{1-v^2}~v_{,V} + \frac{\Omega_F n}{4\pi\eta^2 hu_t} \big[(g_{tt}u_\varphi -\Omega_Fg_{\varphi\varphi}u_t) - V(1-V^2)(g_{tt}u_\varphi -\Omega_Fg_{\varphi\varphi}u_t)_{,V}\big]\,.
\end{equation}
This is the most general equation for $a^2$ 
%ever derived 
in the case of an axisymmetric MHD 
%spherical
accretion onto a spherical BH, it is an implicit equation for the sound speed as its rhs includes $a^2$ too via $u_{t,V}$ and $u_{\varphi,V}$. This equation determines the sound speed at the CP and Eq.~\eqref{CP1} determines the location of the CP once an equation of state (EoS) is supplied.\vskip4pt

We see clearly from ~\eqref{CP3} that at the CP: $v^2\neq a^2\neq V^2$ in general. Even in the special case $\Omega_F=0$ and/or $\eta=\infty$ (no magnetic field if $\Omega_F =0$ and $\eta =\infty$),
\begin{equation}\label{CP4}
	a^2=\frac{vV(1-V^2)}{1-v^2}~v_{,V}\,,\qquad (\Omega_F=0 \text{ and/or } \eta=\infty)\,,
\end{equation}
is still different from $V^2$ and $v^2$, as can be checked from the expression of $v(r,V)$ in these special cases [which can be obtained from the general expressions~\eqref{A1} and~\eqref{A2}]:
\begin{align}\label{CP5}
	&v(r,V)^2 = \frac{16 \pi ^2 (1-V^2) \mathcal{L}^2 \eta ^4+V^2 g_{\varphi \varphi } (4 \pi  \eta ^2 h+g_{t t} n)^2}{16 \pi ^2 (1-V^2)
		\mathcal{L}^2 \eta ^4+g_{\varphi \varphi } (4 \pi  \eta ^2 h+g_{t t} n)^2}\,,\qquad  (\Omega_F =0)\,,\\
	&v(r,V)^2 = \frac{(1-V^2) \mathcal{L}^2+V^2 g_{\varphi \varphi } h^2}{(1-V^2) \mathcal{L}^2+g_{\varphi \varphi } h^2}\,,\qquad  (\eta =\infty)\,,
\end{align}
where ($n,\,h$) are functions of ($r,\,V$) given by~\eqref{k9} and~\eqref{w3} [see also~\eqref{ErV}]. However, if the angular momentum $\mathcal{L}=0$, these last equations reduce to $v^2=V^2$ and~\eqref{CP4} yields $a^2=V^2=v^2$.\vskip4pt

Another way, much simpler, to determine the CPs is to use the variables ($r,\,n$) instead of ($r,\,V$), as we did earlier in~\cite{Azreg2016}. Using $u^\alpha u_\alpha=-1$, \eqref{C1} and~\eqref{L}, we obtain
\begin{align}
	\label{cpn1}&u_t(r,n)=-\frac{8  \pi  \mathcal{L} \eta ^2 \Omega _F^2 g_{t t} g_{\varphi \varphi }n^2+[n g_{t t}+4 \pi  \eta ^2 h(n)] \sqrt{F(r,n)}}{2
		\Omega _F g_{\varphi \varphi } G(r,n)n}\,,\\
	\label{cpn2}&u_{\varphi }(r,n)=\frac{8 \pi  \mathcal{L} \eta ^2 [n g_{t t}+4 \pi  \eta ^2 h(n)]-\sqrt{F(r,n)}}{2 G(r,n)}\,,\\
	&G(r,n)=g_{t t}^2 n^2+16 \pi ^2 \eta ^4 h(n)^2+g_{t t} [\Omega _F^2 g_{\varphi \varphi }n+8 \pi  \eta ^2 h(n)] n\,,\\
	&F(r,n)=64 \pi ^2 \mathcal{L}^2 \eta ^4 [n g_{t t}+4 \pi  \eta ^2 h(n)]^2-4 [16 \pi ^2 \mathcal{L}^2 \eta ^4+\Omega _F^2 (n^2
	g_{t t} g_{\varphi \varphi }^2-C^2)] G(r,n)\,.
\end{align}
This transforms~\eqref{E} to the following Hamiltonian expression
\begin{equation}\label{Ern}
	{\mathcal{E}}(r,n)=-\bigg(h(n)+\frac{\Omega _F^2 g_{\varphi \varphi }(r)\,n}{4 \pi  \eta ^2}\bigg) u_t(r,n)+\frac{\Omega _F g_{t t}(r)\,n}{4 \pi  \eta ^2}u_{\varphi }(r,n)\,.
\end{equation}
Upon differentiating~\eqref{Ern}, we obtain
\begin{equation}\label{w6}
	d{\cal{E}}(r,n)={\cal{E}}_{,r}\,dr + {\cal{E}}_{,n}\,dn =0,
\end{equation} 
since ${\cal{E}}$ is global constant of motion. Thus,
\begin{equation}\label{w7}
	\frac{dn}{dr}=-\frac{{\cal{E}}_{,r}}{{\cal{E}}_{,n}}\,.
\end{equation}
At the CPs where $(r,\,n)=(r_c,\,n_c)$ we have
\begin{equation}\label{CPrn}
	{\cal{E}}_{,r}(r_c,n_c)=0\quad\text{and}\quad {\cal{E}}_{,n}(r_c,n_c)=0\,.
\end{equation}
The explicit forms of these two equations are derived from~\eqref{CP1} and~\eqref{CP2} upon making simple substitutions and noticing that $r$ and $n$ are treated as independent variables
\begin{align}
	\label{CC1}&0=(\ln u_t)_{,r} + \frac{\Omega_F n}{4\pi\eta^2 hu_t}~(\Omega_Fg_{\varphi\varphi}u_t-g_{tt}u_\varphi)_{,r} \,,\\
	\label{CC2}&a^2=-n(\ln u_t)_{,n} + \frac{\Omega_F n}{4\pi\eta^2 hu_t}\bigg[(g_{tt}u_\varphi -\Omega_Fg_{\varphi\varphi}u_t)-n(\Omega_Fg_{\varphi\varphi}u_t-g_{tt}u_\varphi)_{,n}\bigg] \,,
\end{align}
where the partial derivatives with respect to $r$ and $n$ are evaluated using~\eqref{cpn1} and~\eqref{cpn2}. Equation~\eqref{CC2} is another general implicit equation for $a^2$. 
%ever derived in MHD spherical accretion.
\vskip4pt

\subsection{Polytropic Fluid}

If the equation of state is that of a polytropic fluid
%\begin{equation}
%	p=S(m n)^{1+\frac{1}{N}}=Kn^\gamma,\qquad (\gamma=(N+1)/N,\;K:=Sm^\gamma),
%\end{equation}
\begin{equation}\label{pressure}
	p=Kn^\gamma,\quad \text{where } \gamma=(N+1)/N\,.
\end{equation}
Here $K$ is a constant, $m$ is the baryonic mass of the plasma particle, and $N$ is the polytrope index. Following~\cite{Azreg2016} we can show that the specific entropy satisfies
\begin{equation}\label{entropy}
u^\mu\nabla_\mu s\propto u_\mu F^{\mu\nu}J_\nu \,.
\end{equation}
Here $J_\nu\propto u_\nu$ is the current density 4-vector. In case when the ideal MHD flow condition~\eqref{IdealMHDcond} holds, we have $u^\mu\nabla_\mu s=\partial_r s= 0$, implying an isentropic flow and in this case the specific enthalpy and the three-dimensional sound speed are given by, respectively~\cite{Azreg2016,Azreg2017}
\begin{equation}\label{w3}
	h(r,V) = m\Big(1+\frac{k\gamma n^{\gamma-1}(r,V)}{\gamma-1}\Big),\qquad (K=mk)
\end{equation}
\begin{equation}\label{w4}
	a^2(r,V)=\frac{k\gamma(\gamma-1)n^{\gamma-1}}{(\gamma-1)+k\gamma n^{\gamma-1}},
\end{equation}
%and
%\begin{equation}\label{MA}
%	M_A^2(r,V)=4\pi \eta^2 m \Big[ n^{-1}(r,V)+\frac{k\gamma n^{\gamma-2}(r,V)}{\gamma-1}\Big]=4\pi \eta^2\, \frac{K\gamma n^{\gamma-2}(r,V)}{a^2(r,V)}\,,
%\end{equation}
where $n(r,V)$ is given in~\eqref{k9}. It is straightforward to show that
\begin{equation}\label{w5}
	h=\frac{K\gamma n^{\gamma-1}}{a^2}=\frac{m(\gamma-1)}{(\gamma-1)-a^2}\,,
\end{equation}
implying $a^2<\gamma-1$.

\begin{figure}[H]
	\centering
	\includegraphics[width=0.325\linewidth]{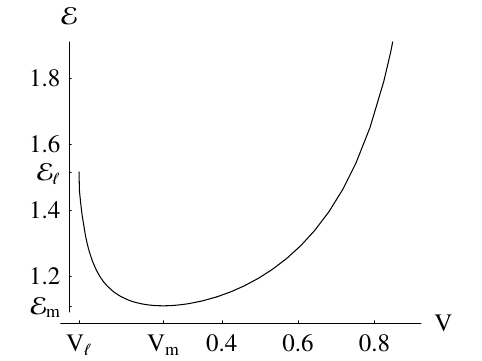}
	\includegraphics[width=0.325\linewidth]{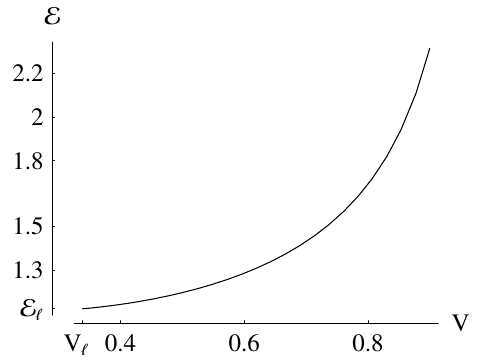}
	\includegraphics[width=0.325\linewidth]{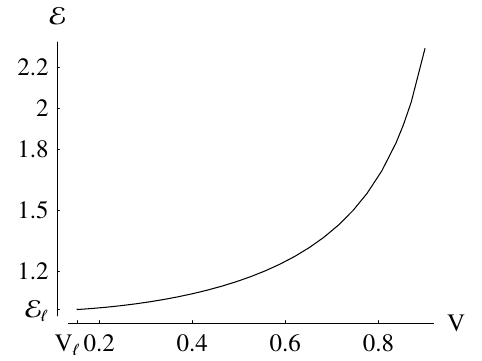}
	\caption{\footnotesize{$\mathcal{E}/m_g$ ($\mathcal{E}$ in the units of $m_g$) versus $V$ upon fixing $r$ at $r_{\text{out}}$ and varying the intensity of the magnetic field $\Phi_1$~\eqref{constants}. Left panel: $r_{\text{out}}=100r_g$ and $y=8$. Below the speed $V_\ell =0.02581$ the in/outflow is not possible. Above ${\mathcal{E}}_\ell =1.51627$, there is only one possible speed of the flow. Below ${\mathcal{E}}_\ell$ and above ${\mathcal{E}}_{\text{m}} =1.11166$ (corresponding to $V_{\text{m}} =0.24663$), there are two possible initial speeds of the flow (relativistic and non-relativistic).  Intermediate panel: $r_{\text{out}}=100r_g$ and $y=7.5$. Upon increasing the the intensity of the magnetic field $\Phi_1$, $V_\ell =0.339$ increases and the minimum of ${\mathcal{E}}_\ell$ disappears. No possible flow occurs below ${\mathcal{E}}_\ell =1.12535$ and above it only relativistic flow is possible. Right panel: $r_{\text{out}}=1000r_g$ and higher magnetic field $y=6.7$. In this case $V_\ell =0.15224$ (for the corresponding case $r_{\text{out}}=100r_g$ and $y=6.7$, $V_\ell =0.99786$ is extremely relativistic). No possible flow happens below ${\mathcal{E}}_\ell =1.01760$.}}
	\label{FigeV}
\end{figure}

\section{Special features in MHD Schwarzschild spherical accretion}
\subsection{Global parameter values}

In this work $m$ is supposed to be proton mass $1.67262192\times 10^{-27}$ kg ($m_g=1.24187391\times 10^{-54}$ in geometrized units). Unless otherwise stated, the other constants in geometrized units are as follows:
\begin{align}\label{constants}
	&k=5.08445\times 10^{-17}\,,\quad C=\frac{7\times 10^{12}\times (M/M_\odot)}{12\pi\times m (\text{kg})\times c (\text{m/s})}\,,\quad N=\frac{4}{3}\,,\quad \gamma =\frac{7}{4}\,,\nonumber\\ &{\mathcal{L}}=3.10m_g\,,\quad\Omega_F =0\,,\quad \Phi_1 =-7\times 10^{-y}\,,\quad \eta =-\frac{C}{\Phi_1}\,,
\end{align}
where $M$ and $M_\odot$ are the masses of the BH and the sun, respectively, and $c$ is the speed of light. The value of $C$ corresponds to $0.001$ times the Eddington rate, which is $1.44\times 10^{15}$ (kg/s) taking the average ratio $r/H$, where $H$ is the height of the disk, to be $0.6$. To show the main feature of MHD flow we mainly focus on the case $\Omega_F =0$ unless otherwise stated.
Note that if a flow solution passes thorough the \alf point in $(r,V)$-plane or $(r,n)$-plane, where the denominator of \eqref{k5} becomes zero, the regularity condition \eqref{k5} yields a constraint between $\Omega_F$ and ${\cal{L}}$ that does not allow the case $\Omega_F=0, {\cal{L}}\neq 0$ as discussed in \cite{Takahashi1990,Takahashi2006,Pu2015}. In this paper, we focus on the solutions that do not cross the \alf point which happens when 
$\eta$ is large or correspondingly magnetic field is weak. Therefore, $\Omega_F$ and ${\cal{L}}$ are independent of each other.
For the case of the Schwarzschild BH, the MHD accretion was intensively studied in~\cite{Mitra:2022iiv} and other references therein. In this work we aim to show those special features of flow not yet investigated in the scientific literature.\vskip4pt

\subsection{Results}
As an initial condition we assume that the outer boundary of the flow (of the disk) is at $r_{\text{out}}=100r_g$ with $r_g=GM/c^2$ and that $M=M_\odot$. We have also examined the case with $r_{\text{out}}=1000r_g$ and the main features remain absolutely the same with the only difference that the lower possible initial speed $V_\ell$ of the in/outflow at $r_{\text{out}}$ depends on $r_{\text{out}}$ itself and on the other parameters. For $\Omega_F=0$, the expression of $u_\varphi$~\eqref{k5} is positive if its denominator is positive. So, the lower speed $V_\ell$ at $r_{\text{out}}$ is solution to $4\pi \eta^2 h +g_{tt} n=0$. For fixed $r_{\text{out}}$, the lower speed $V_\ell$ is an increasing function of the intensity of the magnetic field $\Phi_1$~\eqref{constants}, as shown in Fig.~\ref{FigeV}. In this figure we plot $\mathcal{E}/m_g$ ($\mathcal{E}$ in the units of $m_g$) versus $V$ upon fixing $r$ at $r_{\text{out}}$ and varying the intensity of the magnetic field $\Phi_1$~\eqref{constants}. Left panel: $r_{\text{out}}=100r_g$ and $y=8$. Below the speed $V_\ell =0.02581$ the in/outflow is not possible. Above ${\mathcal{E}}_\ell =1.51627$, there is only one possible speed of the flow. Below ${\mathcal{E}}_\ell$ and above ${\mathcal{E}}_{\text{m}} =0.24663$, there are two possible initial speeds of the flow (relativistic and non-relativistic).  Intermediate panel: $r_{\text{out}}=100r_g$ and $y=7.5$. Upon increasing the intensity of the magnetic field $\Phi_1$, $V_\ell =0.339$ increases and the minimum of ${\mathcal{E}}_\ell$ disappears. No possible flow below ${\mathcal{E}}_\ell =1.12535$ and above it only relativistic flow is possible. Right panel: $r_{\text{out}}=1000r_g$ and higher magnetic field $y=6.7$. In this case $V_\ell =0.15224$ (for the corresponding case $r_{\text{out}}=100r_g$ and $y=6.7$, $V_\ell =0.99786$ is extremely relativistic). No possible flow occurs below ${\mathcal{E}}_\ell =1.01760$. As $\Phi_1$ increases, the magnetic pressure increases too preventing the flow from advancing unless its initial speed exceeds some lower limit $V_\ell$ at the outer boundary $r_{\text{out}}$ of the disk.\vskip4pt

\subsubsection{Outflow with constant limiting speed - Inflow in the vicinity of the horizon}
\begin{figure}[H]
	\centering
	\includegraphics[width=0.44\linewidth]{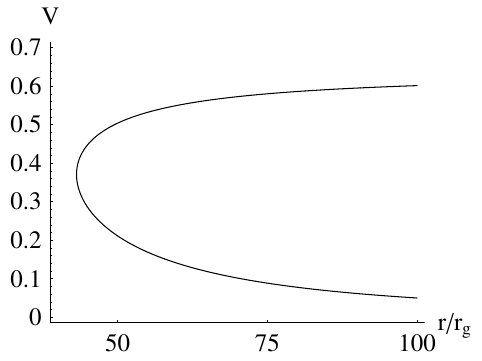}
	\includegraphics[width=0.44\linewidth]{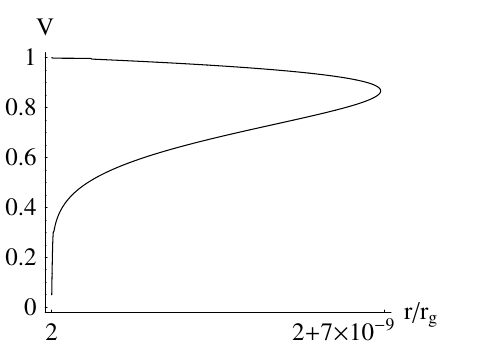}
	\caption{\footnotesize{$V$ versus $r/r_g$ for $\mathcal{E}=\mathcal{E}(100r_g,0.05)=1.28736$ ($\mathcal{E}/m_g=1.28736$) and $y=8$. Left panel: Outflow with constant limiting speed. If the fluid is ejected at $r=43.1659r_g$ with a speed of $V_i=0.371$, it may flow along the upper branch and it reaches the final speed $V_f=0.629765$ at spatial infinity (Note that the speed of the fluid as it reaches $100r_g$ is $0.629765$, as can also be seen from the left panel of Fig.~\ref{FigeV}). The fluid may flow along the lower path and it reaches anther nearly zero speed at spatial infinity (Note that the speed of the fluid as it reaches $100r_g$ is $0.05$, as can also be seen from the left panel of Fig.~\ref{FigeV}). Right panel: Inflow. In the vicinity of the horizon, the level curve $\mathcal{E}=\mathcal{E}(100r_g,0.05)=1.28736$ has another branch, where the fluid may flow along the upper branch and it crosses the horizon with a speed $V=1$, or flow along the lower branch and crosses the horizon with a speed $V=0.0387543$. In this region the magnetic field is so strong that the fluid has to start its journey with a speed nearing $0.85$ with $r$ being around $2+7\times 10^{-9}$.}}
	\label{FigVr8a}
\end{figure}

In Fig.~\ref{FigVr8a}, we plot $V$ versus $r/r_g$ for $\mathcal{E}=\mathcal{E}(100r_g,0.05)=1.28736$ and $y=8$ with conditions at the outer boundary $r_{\text{out}}=100r_g$ and $V_{\text{out}}=0.05$. The left panel depicts the outflow. If the fluid is ejected at $r=43.1659r_g$ with a speed of $V_i=0.371$, it may flow along the upper branch and it reaches the final speed $V_f=0.629765$ at spatial infinity, as can be seen from the Taylor expansion
\begin{multline}
	\mathcal{E}(r,V)=1.28736-\frac{1900.95}{r}+\Big(1.34362-\frac{1984.02}{r}\Big) (V-0.629765)\\
	+\Big(3.17027-\frac{4681.3}{r}\Big) (V-0.629765)^2+\cdots
\end{multline} 
The fluid may flow along the lower path and it reaches a zero speed at spatial infinity; more precisely, the speed at $r=10^{100+q}$ and $q>0$ is $V=5.22082\times 10^{-198-2q}$. The right panel depicts the inflow. In the vicinity of the horizon, the level curve $\mathcal{E}=\mathcal{E}(100r_g,0.05)=1.28736$ has another physical branch, where the fluid may flow along the upper branch and it crosses the horizon with a speed $V=1$, or flow along the lower branch and crosses the horizon with a speed $V=0.0387543$. In this region the magnetic field is so strong that the fluid has to start its journey with a speed nearing $0.85$ with $r$ being around $2+7\times 10^{-9}$.\vskip4pt

\subsubsection{Non-critical global flow}
\begin{figure}[H]
	\centering
	\includegraphics[width=0.45\linewidth]{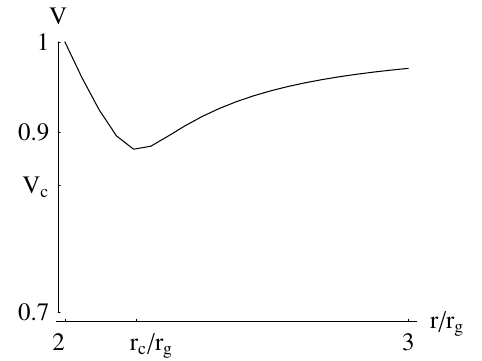}
	\includegraphics[width=0.45\linewidth]{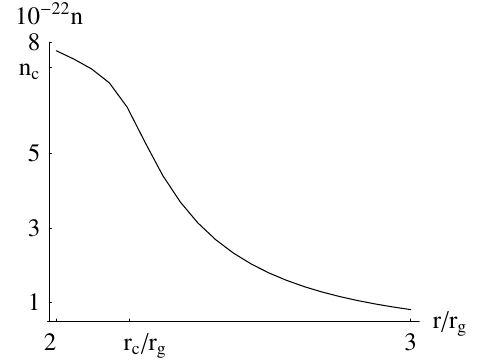}
	\includegraphics[width=0.45\linewidth]{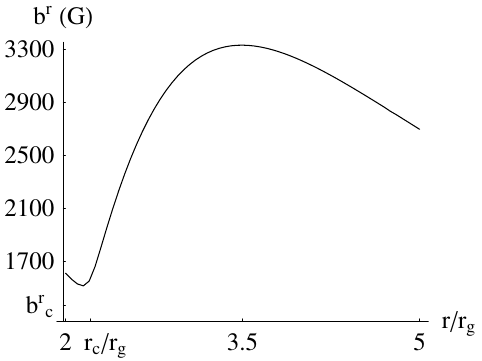}
	\includegraphics[width=0.45\linewidth]{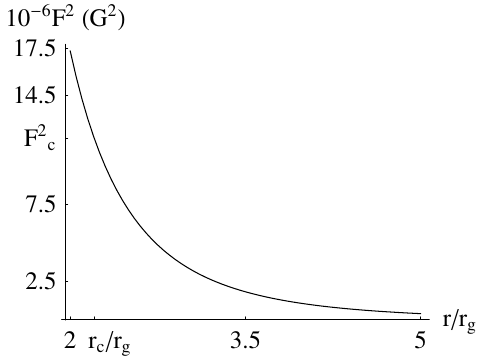}
	\caption{\footnotesize{Non-critical global flow. Plots of $V$, $n$, $b^r$, and $F^2$ versus $r/r_g$ for $\mathcal{E}=10.08736$ and $y=8$ with the outer boundary $r_{\text{out}}=100r_g$. In each panel the plot passes near the CP [defined by ${\cal{E}}_{,r}(r_c,V_c)=0$ and ${\cal{E}}_{,V}(r_c,V_c)=0$] but not through it. At the CP we have $r_c=3258.28=2.20657r_g$, $V_c=0.841218$, $n_c=7.32727\times 10^{22}$, $b^r{}_c=1369.31\text{ G}$, $F^2{}_c=1.1711\times 10^7\text{ G}^2$. At the outer boundary of the disk we have $r_{\text{out}}=100r_g$, $V_{\text{out}}=0.995119$, $n_{\text{out}}=1.70121\times 10^{18}$, $b^r{}_{\text{out}}=11.8193\text{ G}$, $F^2{}_{\text{out}}=2.77628\text{ G}^2$. Here $n_c$, $b^r{}_c$, and $F^2{}_c$ are the value of  $n(r,V)$, $b^r(r,V)$, and $F^2(r,V)$ at the point $(r,V)=(r_c,V_c)$.}}	
	\label{FigVr8b}
\end{figure}

Keeping the same condition at the outer boundary $r_{\text{out}}=100r_g$ and $y=8$, and increasing $\mathcal{E}$ to $\mathcal{E}=10.0874$, the flow becomes global and non-critical, as shown in Fig.~\ref{FigVr8b}. In each of the four panels, the plot passes near the CP [defined by ${\cal{E}}_{,r}(r_c,V_c)=0$ and ${\cal{E}}_{,V}(r_c,V_c)=0$] or by ${\cal{E}}_{,r}(r_c,n_c)=0$ and ${\cal{E}}_{,n}(r_c,n_c)=0$) but not through it. Based on the left panel of Fig.~\ref{FigeV}, the level curve of $\mathcal{E}=10.08736$ in the $rV$-plane, which mathematically has two branches, physically speaking, only the upper branch (relativistic branch) is relevant. This is because for the lower non-relativistic branch the equation $\mathcal{E}=10.08736=\mathcal{E}(100 r_g,V)$ has a solution $V=0.00051953<V_\ell =0.02581$: The magnetic pressure at the outer boundary $100r_g$ is such that flow with speed below $V_\ell$ is allowed. We see that the radial component of magnetic field, $b^r$, jumps high from its outer value $b^r{}_{\text{out}}=11.8193\text{ G}$ to about $3300\text{ G}$, then drops to about half that value at the horizon $r_{\text{h}}=2r_g$. As to the radial 3-speed $V$, it decreases from $V_{\text{out}}=0.995119$ to $V(r=r_c)=0.880467>V_c=0.841218$, then increases to its final value $V(r=r_{\text{h}})=1^-$. Such global accretion flow starting at $r_{\text{out}}=100r_g$ with such high speed is possible if, for instance, the fluid was ejected (an outflow) by a companion star.\vskip4pt

\subsubsection{Critical global flow}
\begin{figure}[H]
	\centering
	\includegraphics[width=0.45\linewidth]{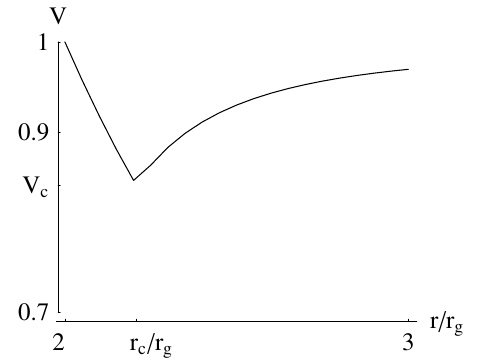}
	\includegraphics[width=0.45\linewidth]{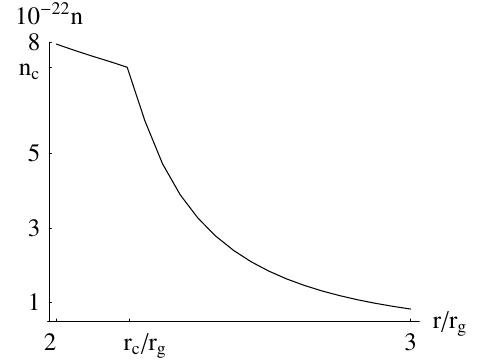}
	\includegraphics[width=0.45\linewidth]{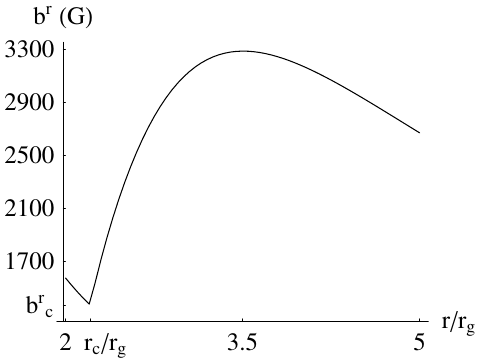}
	\includegraphics[width=0.45\linewidth]{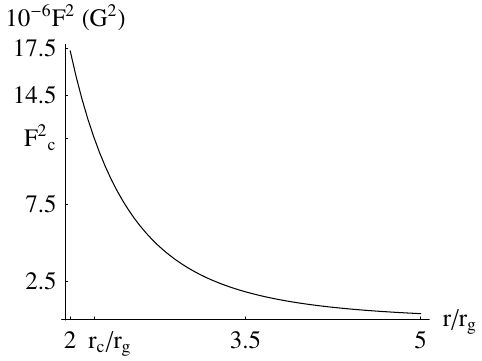}
	\caption{\footnotesize{Critical global flow. Plots of $V$, $n$, $b^r$, and $F^2$ versus $r/r_g$ for $\mathcal{E}=10.02082$ and $y=8$ with the outer boundary $r_{\text{out}}=100r_g$. In each panel the plot passes through the corresponding CP. At the CP we have $r_c=3258.28=2.20657r_g$, $V_c=0.841218$, $n_c=7.32727\times 10^{22}$, $b^r{}_c=1369.31\text{ G}$, $F^2{}_c=1.1711\times 10^7\text{ G}^2$. At the outer boundary of the disk we have $r_{\text{out}}=100r_g$, $V_{\text{out}}=0.995053$, $n_{\text{out}}=1.71276\times 10^{18}$, $b^r{}_{\text{out}}=11.7404\text{ G}$, $F^2{}_{\text{out}}=2.77628\text{ G}^2$.}}	
	\label{FigVr8c}
\end{figure}

A global critical flow is observed for $r_{\text{out}}=100r_g$ with $y=8$ if we take $\mathcal{E}=10.02082$, as shown in Fig.~\ref{FigVr8c}, where each plot passes through the corresponding CP. In this case too, only the upper relativistic branch $V(r)$ is relevant. This case shows a special feature compared to the case of Fig.~\ref{FigVr8b}; each plot is continuous at the CP but a corner appears there indicating a non-smoothness in the slope [exception: the slope of $F^2(r)$ appears to be continuous at the CP]. The non-smoothness in the slope is not an isolated case; rather, we noticed such a behavior for other, even non-critical, values of $\mathcal{E}$, as $10.02126$.\vskip4pt

\subsubsection{Magnetically-arrested-disk state}
\begin{figure}[H]
	\centering
	\includegraphics[width=0.45\linewidth]{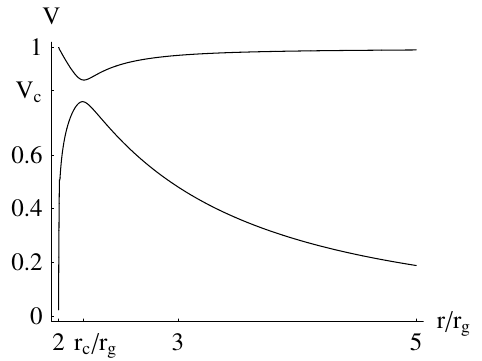}
	\includegraphics[width=0.45\linewidth]{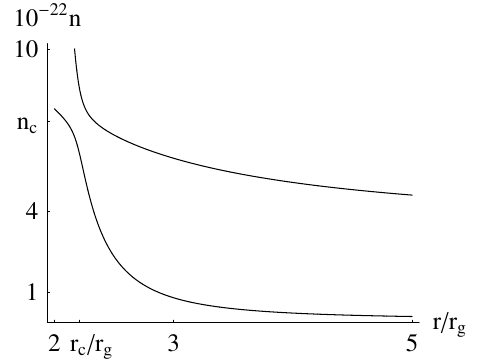}
	\includegraphics[width=0.45\linewidth]{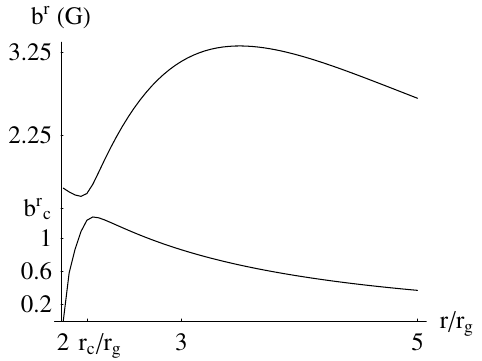}
	\includegraphics[width=0.45\linewidth]{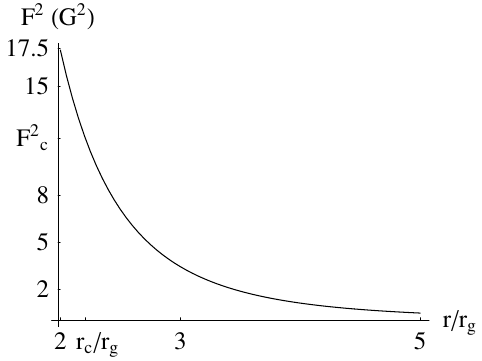}
	\caption{\footnotesize{Magnetically-arrested-disk, non-critical, global flow. Plots of $V$, $n$, $b^r$, and $F^2$ versus $r/r_g$ for $\mathcal{E}=\mathcal{E}(100r_g,0.00052)=10.08127$ and $y=11$ with the outer boundary $r_{\text{out}}=100r_g$. This case corresponds to $V_{\text{out}}=0.00052$, $n_{\text{out}}=3.29906\times 10^{22}$, $b^r{}_{\text{out}}=0.00116635\text{ G}$, $F^2{}_{\text{out}}=2.77632\times 10^{-6}\text{ G}^2$ for the non-relativistic branch, and to $V_{\text{out}}=0.995113$, $n_{\text{out}}=1.70226\times 10^{18}$, $b^r{}_{\text{out}}=0.011812\text{ G}$, $F^2{}_{\text{out}}=2.77628\times 10^{-6}\text{ G}^2$ for the relativistic branch.  At the CP we have $r_c=3258.28=2.20657r_g$, $V_c=0.841217$, $n_c=7.32733\times 10^{22}$, $b^r{}_c=1.36931\text{ G}$, $F^2{}_c=11.711\text{ G}^2$. The upper branch in the $rn$-plane is the non-relativistic one and the lower branch is the relativistic one. The two branches in the $rb^r$-plane are in one-to-one correspondence with the branches in $rV$-plane. The two branches in the $rF^2$-plane coincide.}}	
	\label{FigVr11}
\end{figure}

In order to observe new phenomena, we decrease the intensity of the magnetic field taking $y=11$ and keeping $r_{\text{out}}=100r_g$. All fluid motions met in the case $y=8$ are also met in this case $y=11$ and plus. For this case, a similar curve to the left panel of Fig.~\ref{FigeV} exists, however, it extends up to infinity in both limits $V\to 0^+$ and $V\to 1^-$. This means that for any\footnote{The actual value of $\mathcal{E}_{\text{m}}$ for this case is slightly different from that given in the caption of Fig.~\ref{FigeV}. The values are $\mathcal{E}_{\text{m}}=1.111661501620652$ for $y=8$ and $\mathcal{E}_{\text{m}}=1.1116615015528608$ for $y=11$. The values of $V_{\text{m}}$ are the same.} $\mathcal{E}>\mathcal{E}_{\text{m}}=1.11166$, there are always two branches in the $rV$-plane, non-relativistic and relativistic, for any initial speed $V_{\text{out}}$ of the flow (in or out) at $r_{\text{out}}=100r_g$. We consider the magnetically-arrested-disk, non-critical, state $\mathcal{E}=\mathcal{E}(100r_g,0.00052)=10.08127$ corresponding to $V_{\text{out}}=0.00052$, $n_{\text{out}}=3.29906\times 10^{22}$, $b^r{}_{\text{out}}=0.00116635\text{ G}$, $F^2{}_{\text{out}}=2.77632\times 10^{-6}\text{ G}^2$ for the non-relativistic branch, and to $V_{\text{out}}=0.995113$, $n_{\text{out}}=1.70226\times 10^{18}$, $b^r{}_{\text{out}}=0.011812\text{ G}$, $F^2{}_{\text{out}}=2.77628\times 10^{-6}\text{ G}^2$ for the relativistic branch. The corresponding plots are depicted in Fig.~\ref{FigVr11}. For the non-relativistic branch in the $rV$-plane, the fluid crosses the horizon at $r=r_{\text{h}}$ with a speed $V_{\text{h}}=0.00249148$ and, because of the conservation law~\eqref{k9}, $n$ diverges at $r=r_{\text{h}}$ where $g_{tt}(r_{\text{h}})=0$. The sharp decrease of the speed in the non-relativistic branch $V(r)$, from about $V_c$ to $V_{\text{h}}$ within an interval of extent $\Delta r=0.20657r_g$, accompanied by a similar decrease in $b^r$ and a sharp increase in $n$, is a signal of the magnetically-arrested-disk state in the corona $\Delta r$ of high density ($n\to\infty$ as $r\to r_{\text{h}}$) around the horizon. We have obtained similar results for $y=9.5$ for fixed $\mathcal{E}$ and the sharp decrease in $V(r)$ occurs at the same location.\vskip4pt 
%For $\mathcal{E}$ and $\mathcal{L}$ of the order of 1 and 2, respectively, it was shown in~\cite{Mitra:2024sqb} that the shock location depends on the intensity of the magnetic field $\Phi$ and it increases with increasing $\Phi$ (decreasing $p$) and that the shock disappears above some limiting value $\Phi_\text{lim}$. We do not know if such a conclusion applies to energies that are 10 times those considered in~\cite{Mitra:2024sqb}. 
%This seems to explain partly why the shock was absent for $p=8$. With that said, this could also be interpreted as the magnetically-arrested-disk state.\vskip4pt

\subsubsection{Shock-induced flow}
According to~\eqref{entropy}, if the ideal MHD flow condition~\eqref{IdealMHDcond} is violated, the isentropic condition is too violated and conversely. Our enthalpy formula~\eqref{w3} is valid under the assumption of isentropic flow. In order to observe a shock-induced flow at some location $r_{\text{sh}}$, we have to envisage a flow where the entropy is constant in the pre-shock flow $s_{\text{pre}}$ and constant in the post-shock flow $s_{\text{post}}$ and has a finite \emph{positive} jump between the two constants at $r_\text{sh}$: 
\begin{equation}\label{shock}
\Delta s\big|_{r_{\text{sh}}}=s_{\text{post}}-s_{\text{pre}}>0\,.	
\end{equation}
The value of $r_{\text{rh}}$ depends on that of $\Delta s$. According to~\eqref{entropy}, a shock state generates a momentarily electric current in the flow.\vskip4pt

We refine some of the parameters in~\eqref{constants} reducing the values of ($k,\,{\mathcal{L}}$), taking $y=8$, and keeping the other parameters unchanged. 
\begin{equation}\label{parameters}
k=5.08445\times 10^{-19}\,,\quad {\mathcal{L}}=1.90m_g\,,\quad \Phi_1 =-7\times 10^{-y}\,,\quad y =8\,.
\end{equation} 

\begin{figure}[H]
	\centering
	\includegraphics[width=0.47\linewidth]{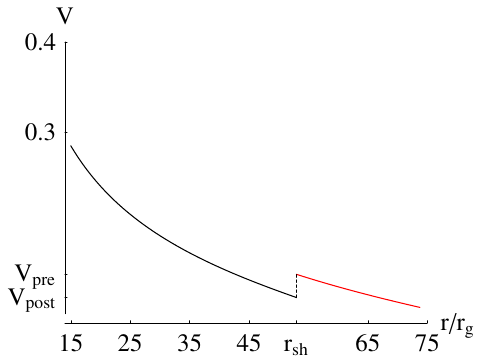}
	\includegraphics[width=0.47\linewidth]{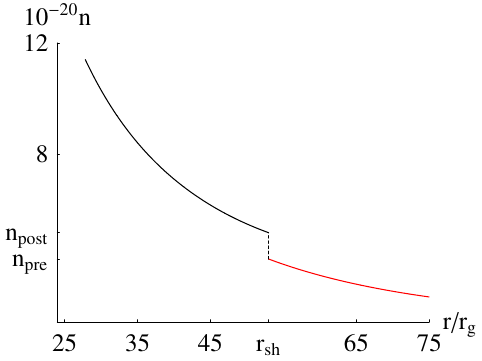}
	\caption{\footnotesize{Shock-induced global flow. Plots of $V$ and $n$ versus $r/r_g$ for ${\mathcal{E}}=0.99468564m_g$, ${\mathcal{L}}=1.90m_g$, and $y=8$ with the other parameters remain unchanged. The pre-shock plots are in red ($k_{\text{pre}}=5.08445\times 10^{-19}$) and the post-shock plots are in black ($k_{\text{post}}=1.7k_{\text{pre}}$). The shock occurs at $r_{\text{sh}}=52.9676r_g$, with $V_{\text{pre}}=0.14353$ and $n_{\text{pre}}=4.25481\times 10^{20}$ [corresponding to points on the red plots (pre-shock state)], and $V_{\text{post}}=0.117889$ and $n_{\text{post}}=5.19795\times 10^{20}$ [corresponding to points on the black plots (post-shock state)]. In the limit $r\to r_{\text{h}}=2r_g$, $n\to 5.20664\times 10^{22}$ and $V\to 1\,$.}}	
	\label{Figshock}
\end{figure}

In shock-induced flow one usually keeps both ${\mathcal{L}}$ and ${\mathcal{E}}$ constant and consider changes in the entropy, number (or rest-mass) density, the 3-speed, and all that that depends of $n$ and $V$ as the magnetic field components~\cite{Mitra:2024sqb,Fukumura:2016guw,Takahashi2006}. Shock-induced flow with energy drop are also possible. In fact, these are the most realistic shock-induced flows and we have obtained instances of such flows taking ${\mathcal{L}}$ as in~\eqref{parameters} and we allowed ${\mathcal{E}}$ to drop by some amount.
%$\Delta {\mathcal{E}}/m_g\big|_{r_{\text{sh}}}=({\mathcal{E}}_{\text{post}}-{\mathcal{E}}_{\text{pre}})/m_g=-0.00044236$ with %${\mathcal{E}}_{\text{post}}=\mathcal{E}(100r_g,0.04)=0.99424328m_g$ and ${\mathcal{E}}_{\text{pre}}=\mathcal{E}(100r_g,0.07)=0.99468564m_g\,$.\vskip4pt
%${\mathcal{E}}_{\text{post}}=0.99424328m_g$ and ${\mathcal{E}}_{\text{pre}}=0.99468564m_g$. 
However, in this work we will only consider flows with no loss of energy. In the following we hold ${\mathcal{L}}$ constant as in~\eqref{parameters} and we keep ${\mathcal{E}}=0.99468564m_g\,$, the values used in Fig.~\ref{Figshock} depicting a non-violent shock. The pre-shock plots are in red and the post-shock plots are in black. Since for polytropic fluids, $k$ is related to the entropy, we had to vary its value from red ($k_{\text{pre}}=5.08445\times 10^{-19}$) to black ($k_{\text{post}}=1.7k_{\text{pre}}$) plots. From pre-shock state to post-shock state, we see a net decrease in the 3-speed and a net increase in the number density at the shock location $r_{\text{sh}}=52.9676r_g$, which is determined numerically upon imposing a discontinuity constraint, Eq.~(10) of~\cite{Fukumura:2016guw} (see also~\cite{Takahashi2006}). 
%given by 
%\begin{equation}\label{shock2}
%	\Big(\frac{h}{n}\Big)_{\text{post}} - \Big(\frac{h}{n}\Big)_{\text{pre}} = -\frac{g}{C^2}~(P_{\text{post}}-P_{\text{pre}}),
%\end{equation}
%where $P=p+F^2/(16\pi)$ and $F^2$ and $p$ are given by~\eqref{n2F2} and~\eqref{pressure}, respectively. Numerically, we obtain $r_{\text{sh}}=3.62149r_g$. 
In Fig.~\ref{Figshock} the values of $V_{\text{pre}}=0.14353$ and $n_{\text{pre}}=4.25481\times 10^{20}$ correspond to points on the red plots (pre-shock state); their values at the post-shock state are $V_{\text{post}}=0.117889$ and $n_{\text{post}}=5.19795\times 10^{20}$. The fluid crosses the horizon at $r\to r_{\text{h}}=2r_g$ with $n= 5.20664\times 10^{22}$ and $V\to 1\,$.

%Knowing the jump $\Delta s$ in the entropy, the latter is evaluated upon solving the equation
%\begin{equation}\label{shockn}
%\Delta s \simeq \frac{mk\gamma n_{\text{sh}}^{\gamma -1}}{T_{\text{sh}}}\Big(\frac{1}{n_{\text{sh}}}-\frac{1}{n_{\text{av}}}\Big)>0,
%\end{equation}
%where $T_{\text{sh}}$ and $n_{\text{sh}}$ are the temperature and number density at $r_{\text{sh}}$ along the pre-shock curve (red) and $n_{\text{av}}$ is the average number density along the vertical line $r=r_{\text{sh}}$ (not shown in Fig.~\ref{Figshock}). This expression is derived upon integrating the thermodynamic relation $Tds=dh-(dp/n)$ along the vertical line $r=r_{\text{sh}}$ from the pre-shock point (on the red plot) to the post-shock point (on the black plot) and assuming $T$ to remain constant along the vertical line $r=r_{\text{sh}}$. As seen from Fig.~\ref{Figshock}, $n_{\text{sh}}<n_{\text{av}}$, yielding $\Delta s>0$.  \vskip4pt

The jump in the value of the entropy is evaluated upon integrating the thermodynamic relation $Tds=dh-(dp/n)$ along the dashed vertical line $r=r_{\text{sh}}$ from the pre-shock point (on the red plot) to the post-shock point (on the black plot) and assuming $T=(mk/k_{\text{B}})n^{\gamma  -1}$ for a perfect gas ($k_{\text{B}}$ being Boltzman's constant). We obtain
\begin{equation}\label{entrpyj}
	\Delta s =\frac{k_{\text{B}}}{\gamma -1}~\ln\Big(\frac{k_{\text{post}}}{k_{\text{pre}}}\Big)=\frac{4k_{\text{B}}}{3}~\ln(1.7)>0\,,
\end{equation}
which does not depend on $r_{\text{sh}}$ and depends only on $k_{\text{post}}/k_{\text{pre}}$ and $\gamma$. We have noticed that the shock state disappears if we choose $k_{\text{post}}=2k_{\text{pre}}$, and this may be a sign that the jump in the entropy could have an upper limit beyond which no shock occurs. This upper limit is certainly not universal as it is parameter dependent.  \vskip4pt

It was shown in~\cite{Mitra:2024sqb} that the shock location $r_{\text{sh}}$ depends on the intensity of the magnetic field $|\Phi_1|$ and it increases with increasing $|\Phi_1|$ and that the shock disappears above some limiting value $\Phi_\text{lim}$; that is, higher values of the magnetic field push the shock location outward, not favoring shock states. We confirm this finding as the discontinuity constraint, Eq.~(10) of~\cite{Fukumura:2016guw}, cannot be satisfied above some value of $|\Phi_1|$. For instance, we have obtained $r_{\text{sh}}=8.37699r_g$, $52.9676r_g$, $69.2255r_g$ for $|\Phi_1|=7\times 10^{-8.5}=2.21359\times 10^{-8}$, $7\times 10^{-8}$, $8\times 10^{-8}$, respectively. We have obtained the further result that the discontinuity constraint cannot be satisfied above some value of $k_{\text{post}}/k_{\text{pre}}$, and thus no shock state. As emphasized in~\cite{Mitra:2024sqb}, these shock states are not isolated, that is, we have obtained many of them for sets of the parameters different from those in~\eqref{parameters}. However, our main purpose in this work is not to give a detailed discussion of the shock state; rather, it is to show how to apply the Hamiltonian approach, skipping thus the task of integrating differential equations, as done so far in the scientific literature, and to determine the profile of the MHD accretion, particularly, the existence of the shock state. In the remaining section, we will consider the quantum corrected Schwarzschild BH and investigate the shock states of MHD accretion onto this BH.\vskip4pt

%\section{The Loop Quantum Corrected BH}
%
%We consider the effective LQG-corrected geodesically
%complete Schwarzschild spacetime whose metric coefficients are given by (with metric signature $(-,+,+,+)$): \cite{Liu:2020ola,Liu:2023vfh}
%\begin{eqnarray}
%	g_{tt}(r)&=&- \frac{(r-r_+)(r-r_-)(r+r_*)^2}{r^4+a_0^2},\nonumber\\
%	g_{rr}(r)^{-1}&=&\frac{(r-r_+)(r-r_-)r^4}{(r^4+a_0^2)(r+r_*)^2},\nonumber\\
%	g_{\theta\theta}(r)&=&r^2+\frac{a_0^2}{r^2},
%\end{eqnarray}
%where 
%\begin{equation*}
%	r_+=\frac{2M}{(1+P)^2},\ \ r_-=\frac{2MP^2}{(1+P)^2}, \ \ r_*=\sqrt{r_+r_-}=\frac{2MP}{(1+P)^2},\ \ P=\frac{\sqrt{1+\epsilon^2}-1}{\sqrt{1+\epsilon^2}+1}
%\end{equation*}
%Here $M$ denotes the ADM mass, $P$ denotes the polymeric function and $\epsilon=\gamma\delta\ll1$, with $\gamma$ and $\delta$ being the Immirzi and polymeric parameter respectively. The parameter $a_0$ denotes the minimum area gap of the LQG.

\section{MHD Accretion for the quantum corrected Schwarzschild BH}

Here we consider the quantum corrected Schwarzschild BH \cite{Lewandowski:2022zce,Yang:2024lmj}
\begin{equation}\label{eq:exteriormetric1}
	\begin{aligned}
		d s^2_{\rm MS}=&-f(r)d t^2+f(r)^{-1}d r^2+r^2d\Omega^2,
	\end{aligned}
\end{equation}
where
\begin{eqnarray}
	f(r)=1-\frac{2M}{r}+\frac{\alpha M^2}{r^4}.
\end{eqnarray}
Here the parameter $\alpha$ is defined as $\alpha=16\sqrt{3}\pi\gamma_{\text{Imm}}{}^3\ell_p^2 $ with $\ell_p=\sqrt{G\hbar}$ denotes the Planck length and $\gamma_{\text{Imm}}$ is the Immirzi parameter.  
It is worth noticing  that  the form \eqref{eq:exteriormetric1}  of the metric is valid for
\begin{eqnarray}\label{eq:rb}
	r\ge r_{\text{b}}=\left(\frac{\alpha GM}{2}\right)^{\frac{1}{3}}.
\end{eqnarray}\vskip4pt

For illustration we take $\alpha =0.5,\,3,\,5$. Within the values of the parameters used in Fig.~\ref{Figshock} and in~\eqref{parameters}, we are guaranteed that $r_{\text{b}}<r_{\text{h}}=2r_g\,$. Table~\ref{Tab1} depicts the variation of $\Delta r_{\text{sh}}$ in terms of $\alpha\,$, where $\Delta r_{\text{sh}}=r_{\text{sh-Q}}-r_{\text{sh}}$, $r_{\text{sh-Q}}$ is the quantum-corrected value of the shock location, and $r_{\text{sh}}=52.9676r_g$ is the (classical) value previously determined in Fig.~\ref{Figshock}. We see that $\Delta r_{\text{sh}}$ slightly increases with $\alpha$, which implies that quantum effects do not favor shock states by pushing the shock location outward.\vskip4pt

\begin{table}
\caption{\footnotesize{Quantum effects on the shock location. Here $\Delta r_{\text{sh}}=r_{\text{sh-Q}}-r_{\text{sh}}$, $r_{\text{sh-Q}}$ is the quantum-corrected value of the shock location, and $r_{\text{sh}}=52.9676r_g$ is the (classical) value previously determined in Fig.~\ref{Figshock}.}}
\label{Tab1}
\begin{tabular*}{0.567\textwidth}{|l|l|l|l|}%
\hline
$\alpha$ & $0.5$ & $3$ & $5$ \\
\hline
$\Delta r_{\text{sh}}/r_g$ & $0.492335\times 10^{-10}$ & $2.80259\times 10^{-10}$ & $4.65832\times 10^{-10}$ \\
\hline
\end{tabular*}
\end{table}

\section{Conclusion}

In this article, we have formulated the accretion dynamics of MHD fluid onto a spherical BH using the Hamiltonian framework. Under the assumptions of ideal MHD and utilizing few conservation laws of energy-angular momentum and particle number, we are able to determine a Hamiltonian expression which is used to derive the Hamilton equations and then their solutions. It was presumed that the flow occurred in the equatorial plane and the only non-vanishing component of magnetic field is the azimuthal one. In the background of a Schwarzschild BH, we have found the inflow / outflow conditions depending on the three velocity and the energy parameters. The phenomena of relativistic and non-relativistic flows  as well as no-flow are also observed. We have also examined the critical flows (i.e., the flows whose phase trajectories pass throught the critical point) as well as non-critical flows (i.e., the flows where phase trajectories pass near the critical point but not through CP) for the Schwarzschild BH. In the later scenario, we have also noticed a global flow i.e., fluid flow starting from the outer edge of the disk and ending at the event horizon. We have also noticed that shocks in MHD plasma near the BH cause particle number density in MHD flow to increase while 3-velocity to decrease locally near the shock region.

In future investigations, we would like to develop the Hamiltonian formulation of MHD flow near a Kerr BH and explore various kinds of flows along with different equations of state. In addition, we are interested in the question how shocks in MHD plasma near BH can generate highly magnetized plasma and whether various quantum corrections to BH geometry can have deeper or observable effects in the accretion process.

\subsection*{Acknowledgments}
MJ would like to thank Zhejiang University of Technology, Hangzhou, China for providing hospitality where part of this work was completed. We would also thank Yen Chin Ong for useful discussions on this project.

\bibliography{Final}

\section*{Appendix: Roots of~\eqref{k8}\label{appA}}
\appendix
\def\theequation{A.\arabic{equation}}
\setcounter{equation}{0}
We substitute $u_\varphi$ given in~\eqref{k5} into~\eqref{k8} and we set $x=\sqrt{1-v^2}$, we obtain two roots for $v^2$ related by $v_1^2(\mathcal{L})=v_2^2(-\mathcal{L})$. Since $\mathcal{L}$ can be given any sign, we will work with the solution $v_1^2$. The two solutions are given by
\begin{equation}\label{A1}
	v^2=v_1^2=1-\left(\frac{4 \pi  \mathcal{L} \eta ^2 \Omega _F \sqrt{-g_{t t}} g_{\varphi \varphi } n \left(1-V^2\right)-\sqrt{H_1}}{H_2}\right)^2,
\end{equation}
\begin{equation}\label{A2}
	v^2=v_2^2=1-\left(\frac{4 \pi  \mathcal{L} \eta ^2 \Omega _F \sqrt{-g_{t t}} g_{\varphi \varphi } n \left(1-V^2\right)+\sqrt{H_1}}{H_2}\right)^2,
\end{equation}
where
\begin{equation}\label{Num}
	H_1=\left(1-V^2\right) g_{\varphi \varphi } \left(4 \pi  \eta ^2 h+g_{t t} n\right)^2 \left[16 \pi ^2 \left(1-V^2\right) \mathcal{L}^2 \eta ^4+\Omega
	_F^2 g_{t t} g_{\varphi \varphi }^2 n^2+g_{\varphi \varphi } \left(4 \pi  \eta ^2 h+g_{t t} n\right)^2\right]\,,
\end{equation}
\begin{equation}\label{Denom}
	H_2=16 \pi ^2 \left(1-V^2\right) \mathcal{L}^2 \eta ^4+g_{\varphi \varphi } \left(4 \pi  \eta ^2 h+g_{t t} n\right)^2\,.
\end{equation}

\end{document}